%% file: BL_Lac_synchrotron_peak.tex
\documentclass[twocolumn,floatfix,tighten]{aastex62}

\received{}
\revised{}
\accepted{}
\submitjournal{ApJ}

\usepackage{graphicx}	
\usepackage{amsmath}	
\usepackage{amssymb}	
\usepackage{bm}		
\usepackage{booktabs}
\usepackage{listings}
\usepackage{color}
\usepackage{threeparttable}

\definecolor{dkgreen}{rgb}{0,0.6,0}
\definecolor{gray}{rgb}{0.5,0.5,0.5}
\definecolor{mauve}{rgb}{0.58,0,0.82}

\newcommand{\ep}{$\textit{E}_\textrm{p}$}
\newcommand{\lp}{$\textit{L}_\textrm{p}$}
\newcommand{\bb}{1$/$\textit{b}}
\newcommand{\aaa}{\textit{a}}

\newcommand{\eplp}{$\textit{E}_\textrm{p}$--$\textit{L}_\textrm{p}$}
\newcommand{\epb}{$\textit{E}_\textrm{p}$--(1$/$\textit{b})}

\newcommand{\rs}{$r_\textrm{s}$}
\newcommand{\ps}{$p_\textrm{s}$}
\newcommand{\rxte}{\textit{RXTE}}
\newcommand{\swift}{\textit{Swift}}
\newcommand{\beppo}{\textit{BeppoSAX}}

\newcommand{\xmm}{\textit{XMM-Newton}}

\newcommand{\bigc}{\textit{C}}
\newcommand{\massaro}{\cite{2008A&A...478..395M}}

\newcommand{\bhmass}{$\textit{M}_\textrm{BH}$}
\let\oldAA\AA
\renewcommand{\AA}{\text{\normalfont\oldAA}}

\usepackage[T1]{fontenc}
\usepackage{ae,aecompl}

\usepackage{newtxtext,newtxmath}

\usepackage{xcolor}
\definecolor{alizarin}{rgb}{0.82, 0.1, 0.26}

\shorttitle{X-ray spectral variations of synchrotron peak in BL Lacs}
\shortauthors{Y.J. Wang {\em et al.}}
\begin{document}

\title{X-ray spectral variations of synchrotron peak in BL Lacs}

\author{Yijun~Wang}
\affiliation{CAS Key Laboratory for Research in Galaxies and Cosmology, Department of Astronomy, University of Science and Technology of China, Hefei 230026, China; yuxuan@mail.ustc.edu.cn, xuey@ustc.edu.cn}
\affiliation{School of Astronomy and Space Science, University of Science and Technology of China, Hefei 230026, China}

\author{Shifu~Zhu}
\affiliation{Department of Astronomy \& Astrophysics, The Pennsylvania State University, University Park, PA 16802, USA; S.F.Z.Astro@gmail.com}
\affiliation{Institute for Gravitation and the Cosmos, The Pennsylvania State University, University Park, PA 16802, USA}

\author{Yongquan~Xue}
\affiliation{CAS Key Laboratory for Research in Galaxies and Cosmology, Department of Astronomy, University of Science and Technology of China, Hefei 230026, China; yuxuan@mail.ustc.edu.cn, xuey@ustc.edu.cn}
\affiliation{School of Astronomy and Space Science, University of Science and Technology of China, Hefei 230026, China}

\author{Minfeng~Gu}
\affiliation{Key Laboratory for Research in Galaxies and Cosmology, Shanghai Astronomical Observatory, Chinese Academy of Sciences, 80 Nandan Road, Shanghai 200030, China}

\author{Shanshan~Weng}
\affiliation{Department of Physics and Institute of Theoretical Physics, Nanjing Normal University, Nanjing 210023, China}

\author{Huynh Anh N. Le}
\affiliation{CAS Key Laboratory for Research in Galaxies and Cosmology, Department of Astronomy, University of Science and Technology of China, Hefei 230026, China; yuxuan@mail.ustc.edu.cn, xuey@ustc.edu.cn}
\affiliation{School of Astronomy and Space Science, University of Science and Technology of China, Hefei 230026, China}
%

\begin{abstract}
The spectral energy distribution of blazars around the synchrotron peak can be well described by the log-parabolic model that has three parameters: peak energy (\ep), peak luminosity (\lp) and the curvature parameter (\textit{b}). 
It has been suggested that \ep\ shows relations with \lp\ and \textit{b} in several sources, which can be used to constrain the physical properties of the emitting region and/or acceleration processes of the emitting particles. 
We systematically study the \eplp\ and \epb\ relations for 14 BL Lac objects using the 3--25~keV \rxte/PCA and 0.3--10~keV \swift/XRT data. Most objects (9/14) exhibit positive \eplp\ correlations, three sources show no correlation, and two sources display negative correlations. In addition, most targets (7/14) present no correlation between \ep\ and \bb, five sources pose negative correlations, and two sources demonstrate positive correlations. 1ES~1959+650 displays two different \eplp\ relations in 2002 and 2016. 
We also analyze \eplp\ and \epb\ relations during flares lasting for several days. The \eplp\ relation does not exhibit significant differences between flares, while the \epb\ relation varies from flare to flare. For the total sample, when \lp\ < $\textrm{10}^\textrm{45}\ \textrm{erg}\ \textrm{s}^\textrm{-1}$, there seems to be a positive \eplp\ correlation. 
\lp\ and the slope of \eplp\ relation present an anti-correlation, which indicates that the causes of spectral variations might be different between luminous and faint sources. \ep\ shows a positive correlation with the black hole mass. We discuss the implications of these results.
\end{abstract}


\section{Introduction}
\label{introd}

Blazars, including BL Lac objects and flat-spectrum radio quasars (FSRQs), are one type of radio-loud active galactic nuclei (AGNs), with the direction of one of their relativistic jets nearly aligned with our line of sight \citep{1995PASP..107..803U}. The emission from blazars is dominated by jet emission whose spectral energy distribution (SED) consists of a low-energy hump that peaks from sub-millimeter wavelengths to X-ray energies and a high-energy hump that peaks from hard X-rays up to TeV $\gamma$-rays. 
The low-energy component is thought to be produced by synchrotron emission of relativistic electrons in the jet, while the high-energy component is probably dominated by the emission from the inverse Compton scattering process \citep[i.e., the leptonic scenario; e.g.,][]{1992ApJ...397L...5M,1992A&A...256L..27D} or related to proton emission processes \citep[i.e., the hadronic scenario; e.g.,][]{2000NewA....5..377A}. According to the peak frequency ($\nu_\textrm{p}$) of the low-energy hump, BL Lac objects can be divided into high-energy peaked BL Lac objects (HBLs; $\nu_\textrm{p}>10^{15}$ Hz), intermediate-energy peaked BL Lac objects (IBLs; $10^{14}$~Hz~$<\nu_\textrm{p}\leq10^{15}$ Hz) and low-energy peaked BL Lac objects (LBLs; $\nu_\textrm{p}\leq10^{14}$ Hz; \citealt{2010ApJ...716...30A}). The synchrotron peak of FSRQs is usually located in the regime from the sub-millimeter band to far-infrared band, and even to optical/UV wavelengths.

The SED around the synchrotron peak can be well described by the log-parabolic model \citep[e.g.,][]{1986ApJ...308...78L, 2004A&A...413..489M,2005A&A...433.1163D,2007A&A...467..501T,2014ApJ...788..179C,2016MNRAS.458...56W,2017ApJ...836...83S,2018A&A...619A..93B,2018MNRAS.480.2046G,2018ApJ...859...49P}, which is characterized by peak energy (\ep), peak luminosity ($\textit{L}_\textrm{p}$), and the curvature parameter around the peak (\textit{b}). For the entire blazar populations, there is an apparent anti-correlation between \lp\ and \ep, widely known as the blazar sequence \citep[e.g.,][]{1998MNRAS.299..433F,1998MNRAS.301..451G,2011ApJ...735..108C,2017MNRAS.469..255G}. However, it is suggested that this trend might be due to the selection bias \citep[e.g.,][]{2012MNRAS.420.2899G} or might disappear after applying the Doppler boosting correction \citep[e.g.,][]{2008A&A...488..867N,2009RAA.....9..168W,2014JApA...35..381H, 2017ApJ...835L..38F, 2018ApJ...867...68W}. Blazars exhibit intense variability in all detectable wavelengths, which, for instance, is illustrated by their seemingly scale-invariant X-ray flares that last from years down to days and even to minutes \citep[e.g.,][]{2004ApJ...605..662C,2005ApJ...622..160X,2018ApJ...853...34Z}. During these flaring periods, \ep, \lp, and \textit{b} may change with fluxes. For an individual object, there appears to be an apparent positive correlation between \ep\ and \lp\ \citep[e.g.,][]{2004ApJ...601..759T,2007A&A...466..521T,2008A&A...478..395M,2009A&A...501..879T,2016MNRAS.461L..26K,2017MNRAS.469.1655K,2018MNRAS.473.2542K}, which might be connected with the physical conditions in the emitting region, e.g., the average electron energy, magnetic field and beaming factor \citep[e.g.,][]{2009A&A...501..879T}. It is also suggested that there might be an apparent anti-correlation between \ep\ and \bb, which could be related to the acceleration processes of emitting particles \citep[e.g.,][]{2004A&A...413..489M,2011ApJ...739...66T}. 

However, only a few studies have focused on \eplp\ and \epb\ relations in individual sources and the largest sample of such previous studies only includes five objects \citep{2008A&A...478..395M}. Therefore, in this work, we use the hitherto largest sample of this kind, which includes 14 BL Lac objects, to systematically study the \eplp\ and \epb\ relations for every single source, aiming to provide stringent observational constraints upon the physical properties and acceleration mechanisms of emitting particles during flares for future theoretical studies. 
We combined the data from the \textit{Rossi X-Ray Timing Explorer} (\rxte) and \textit{Neil Gehrels Swift Observatory} (\swift), for the following reasons: (1) generally, both of them had performed more X-ray observations for multiple blazars than other satellites used in the previous studies (see Table \ref{table:table1}); and (2) the combination of their different spectral coverages significantly enlarges the observational energy range reached. 

\indent This paper is organized as follows. In Section \ref{sec:sample}, we describe the criteria to select our sample and relevant observational data reduction. 
In Section \ref{sec:method}, we show the empirical models that are used to fit the X-ray spectra and the method of calculating parameters. In Section \ref{sec:results}, we analyze the \eplp\ and \epb\ relations of each source during flaring periods, and discuss the correlations of the total sample. Finally, we summarize our results in Section \ref{sec:summary}.  We adopt the following flat $\Lambda$CDM cosmological parameters: $\textit{H}_\textrm{0}$=67.8 km s$^{\textrm{-1}}$ Mpc$^{\textrm{-1}}$, $\textrm{$\Omega$}_\textit{m}$=0.308, and $\textrm{$\Omega$}_{\textrm{$\Lambda$}}$=0.692 \citep{2016A&A...594A..13P}.

\section{Sample and Data Reduction}
\label{sec:sample}

\subsection{Sample Construction}
\label{sec:samcon}

Our sample of 14 representative BL Lac objects (hereafter the total sample; see Table~\ref{table:table1}) was built upon the \rxte\ TeV blazar sample of \cite{2018ApJ...867...68W}, which includes 2 FSRQs, 1 LBLs, 5 IBLs and 24 HBLs. 
The total sample was constructed in three steps.
Firstly, we only focused on HBLs because their synchrotron peaks fall into the X-ray bands that are covered by \rxte\ and \swift. 
Secondly, in order to assure reasonable signal-to-noise ratios, we required that the average 3--25 keV flux of each source \citep[presented in Table 1 of][]{2018ApJ...867...68W} is larger than $\textrm{10}^{-\textrm{11}}\ \textrm{erg}\ \textrm{cm}^{-2}\ \textrm{s}^{-1}$ and
the total counts are larger than 200/20 for each \rxte/\swift\ observation. 
Finally, we excluded the sources with a total number of \rxte\ and \swift\ observations less than 15.

\subsection{RXTE Data Reduction }
\label{rxtedata}

\indent \rxte\ carries on board the All-Sky Monitor (ASM; 1.5--12 keV), Proportional Counter Array (PCA; 2--60 keV) and High Energy X-Ray Timing Experiment (HEXTE; 15--250 keV). In order to obtain high-quality spectra, we utilized the 3--25 keV data of PCA that consists of five nearly identical proportional counter units (PCUs). Following \citet{2011ApJS..193....3R}, before 1998 December 23, we extracted spectra from PCUs 0, 1 and 2; from 1998 December 23 to 2000 May 12, the spectra were extracted from PCUs 0 and 2; after 2000 May, because PCUs 1, 3 and 4 had high-voltage breakdown issues and the propane layer of PCU 0 could not operate after 2000 May 12, we only extracted the spectra from PCU 2.

\begin{deluxetable*}{rlclcc}[t]
\tablecaption{Source sample$^a$ \label{table:table1}}
\tablehead{\colhead{\#} & \colhead{Object Name} & \colhead{\emph{z}$^b$} & \colhead{Data Origin$^c$} & \colhead{$\log (\textit{M}_\textrm{BH}/\textit{M}_{\odot})$} & \colhead{Ref.$^d$}  } 
\startdata
\input{table1.tex}
\enddata
\tablenotetext{}{\textbf{N\textsc{otes.}} $^{a}$ All sources are HBLs. $^b$ Redshift, provided by \href{http://tevcat.uchicago.edu/}{http://tevcat.uchicago.edu/}. $^{c}$ Superscript numbers indicate different data origins: 1: \rxte; 2: \swift; 3: \beppo; and 4: \xmm. 
The two numbers in the parentheses denote the numbers of \rxte\ and \swift\ observations used in this work, respectively (also see Figure~\ref{fig:lc}).
Previous studies are showed in the abbreviated forms: G06: \cite{2006ApJ...644..742G}; M08: \massaro; K17: \cite{2017MNRAS.469.1655K}; and K18: \cite{2018MNRAS.473.2542K}. $^d$ References of the black hole mass: Wu09: \cite{2009RAA.....9..168W}; Woo05: \cite{2005ApJ...631..762W}; Wagner08: \cite{2008MNRAS.385..119W}. \cite{2009RAA.....9..168W} estimated the black hole mass using the correlation between the \textit{R}-band absolute magnitude of host galaxy and the black hole mass. \cite{2005ApJ...631..762W} and \cite{2008MNRAS.385..119W} estimated the black hole mass through the measured stellar velocity dispersion of host galaxies.} 
\end{deluxetable*}

\indent We used {\sc{ftools}} (version 6.21) to reduce the data. Firstly, we followed the standard procedure\footnote{See \href{http://heasarc.gsfc.nasa.gov/docs/xte/recipes/cook\_book.html}{http://heasarc.gsfc.nasa.gov/docs/xte/recipes/cook\_book.html} for details.} to create the data filter file and corresponding good time intervals (GTIs) file. Secondly, we used the latest faint and bright background models to simulate background events of low-flux observations (count rates < 40 counts s$^\textrm{--1}$ PCU$^\textrm{--1}$) and high-flux observations (count rates $\ge$ 40 counts s$^\textrm{--1}$ PCU$^\textrm{--1}$), respectively. Finally, we extracted the total spectrum and background spectrum for each observation. We obtained the net source spectrum by subtracting the background contribution from the total spectrum. The net source spectra were binned to ensure at least 20 counts per bin in order to utilize the $\chi^2$ minimization fitting method.

\subsection{Swift Data Reduction}
\label{swiftdata}

We collected the 0.3--10 keV data from the X-ray Telescope (XRT) carried by \swift. To obtain high-quality spectra, we utilized its Photon Counting (PC) mode observations in this work. The data reduction was performed with the XRT Data Analysis Software ({\sc{xrtdas}}; v.2.4) that is a part of the {\sc{heasoft}} package (v.6.21). The cleaned event files were produced using the {\sc{xrtpipeline}} task with standard filtering criteria. The spectra of the source and background were extracted using the {\sc{xselect}} task. Firstly, we extracted the source spectrum from a circular region with a radius of 30 pixels (1 pixel = 2.36 arcseconds). If the source count rate was above 0.5 count s$^{-1}$, the pile-up effect should be considered. To remove this effect, we re-extracted the source spectrum from an annular region with an inner radius of 1--17 pixels and an outer radius of 30 pixels. The inner radius was selected according to the deviation of the observed point spread function (PSF) from the known, un-piled-up PSF. The background events were extracted from a source-free circular region with a radius of 60 pixels. The ancillary response files (ARFs) were created using the {\sc{xrtmkarf}} task after correcting for the pile-up effect. Source spectra were binned to ensure at least 1 count per bin in order to use the Cash-statistics fitting method. For Mrk~501 and 1ES~1959+650, given that most of their \swift\ spectra were analyzed in detail in the previous works \citep[][]{2017MNRAS.469.1655K,2018MNRAS.473.2542K}, we did not perform the data reduction for their \swift\ data in this work.

\begin{deluxetable*}{rlrrrrrrrr}[t]
\tablecaption{Fitting results of \eplp\ and \epb\ relations \label{table:table2}}
\tablehead{\colhead{} & \colhead{} & \multicolumn{4}{c}{\eplp\ relation} & 
           \multicolumn{4}{c}{\epb\ relation} \\
           \cmidrule(l{3pt}r{3pt}){3-6} \cmidrule(l{3pt}r{3pt}){7-10}
           \colhead{} & \colhead{Object} &
           \multicolumn{2}{c}{Spearman} &
           \multicolumn{2}{c}{$\log {\textit{L}_\textrm{p}}=\alpha \log {\textit{E}_\textrm{p}}+\beta$} &
           \multicolumn{2}{c}{Spearman} &
           \multicolumn{2}{c}{$\textrm{1}/\textit{b}=\textit{C} \log {\textit{E}_\textrm{p}}+\textit{D}$} \\
           \cmidrule(l{3pt}r{3pt}){3-4} \cmidrule(l{3pt}r{3pt}){5-6} \cmidrule(l{3pt}r{3pt}){7-8} \cmidrule(l{3pt}r{3pt}){9-10}
           \colhead{} & \colhead{Name} & \colhead{$r_\textrm{s}$} & \colhead{$p_\textrm{s}$} & 
           \colhead{$\alpha$} & \colhead{$\beta$} & \colhead{$r_\textrm{s}$} & \colhead{$p_\textrm{s}$} & \colhead{\textit{C}} & \colhead{\textit{D}} \\
           \colhead{\#} &
           \colhead{(1)}        &
           \colhead{(2)}        &
           \colhead{(3)}        &
           \colhead{(4)}        &
           \colhead{(5)}        &
           \colhead{(6)}        &
           \colhead{(7)}        &
           \colhead{(8)}        &
           \colhead{(9)}        } 
\startdata
\input{table2.tex}
\enddata
\tablenotetext{}{\textbf{N\textsc{otes.}} Column (1): Source name. Columns (2) and (3): Spearman's rank correlation coefficient and probability of the hypothesis test of \eplp\ relation. Columns (4) and (5): Best-fitting $\alpha$ and $\beta$ of $\log {\textit{L}_\textrm{p}}=\alpha \log {\textit{E}_\textrm{p}}+\beta$, where $E_\textrm{p}$ is in units of keV and $L_\textrm{p}$ is in units of $\textrm{10}^\textrm{44}\ \textrm{erg}\ \textrm{s}^{-1}$. Columns (6) and (7): Spearman's rank correlation coefficient and probability of the hypothesis test of \epb\ relation. Columns (8) and (9): Best-fitting \textit{C} and \textit{D} of $\textrm{1}/\textit{b}=\textit{C} \log {\textit{E}_\textrm{p}}+\textit{D}$, where $E_\textrm{p}$ is in units of keV.}
\end{deluxetable*}

\begin{figure*}[t]
\centering
\includegraphics[width=\linewidth, clip]{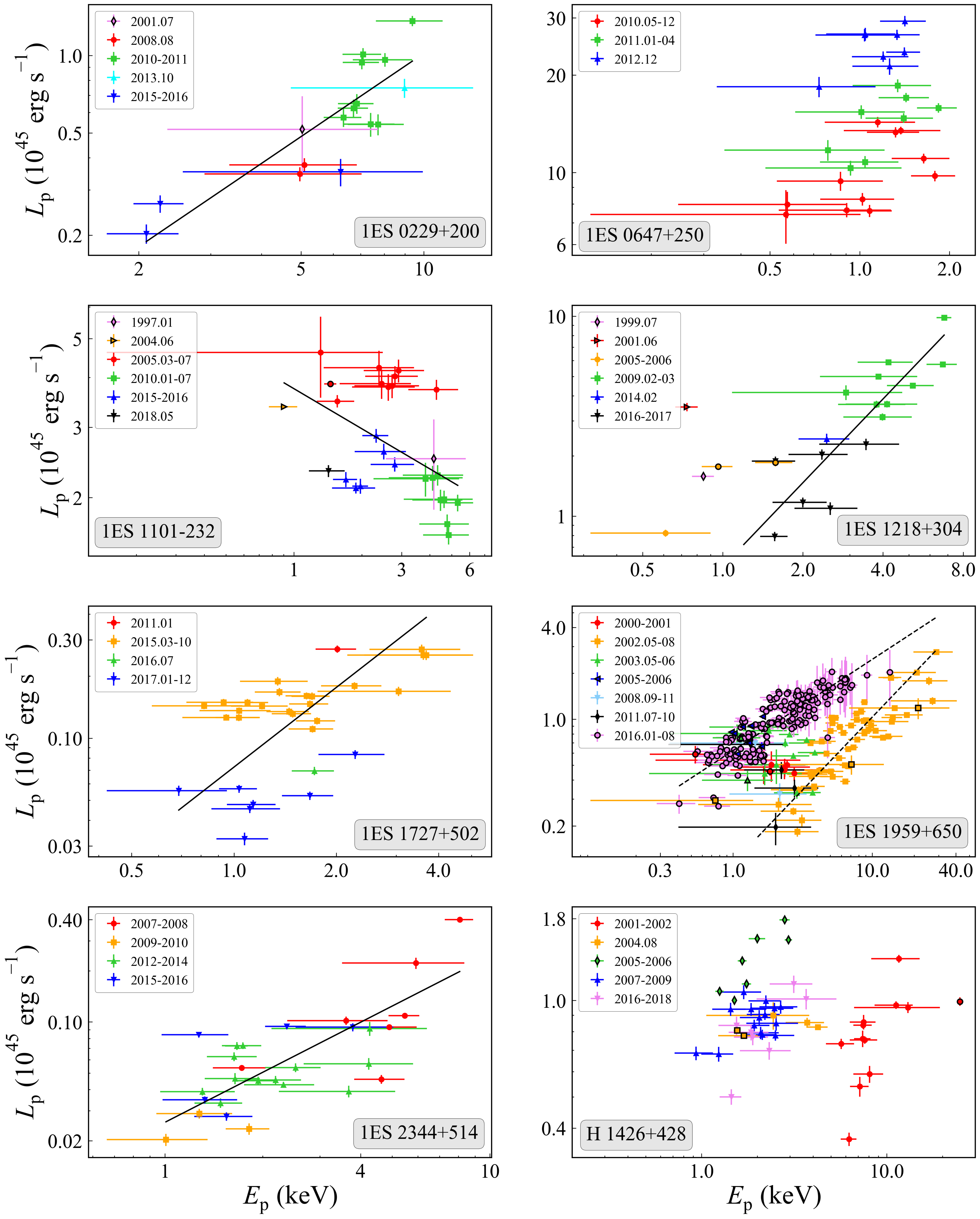}
\caption{\eplp\ relations of the 14 sources in the total sample. The filled circles with black edges represent the literature data, while those without black edges represent the data analyzed in this work (see Table \ref{table:table1}). Different colors indicate different observational dates. The lines denote the best-fitting \eplp\ relations when applicable; for 1ES~1959+650, the two dashed lines are for the 2002 and 2016 data, respectively (see Table~\ref{table:table2}).}
\label{fig:eplp}
\end{figure*}

\renewcommand{\thefigure}{\arabic{figure} (Cont.)}
\addtocounter{figure}{-1}

\begin{figure*}[t]
\centering
\includegraphics[width=\linewidth, clip]{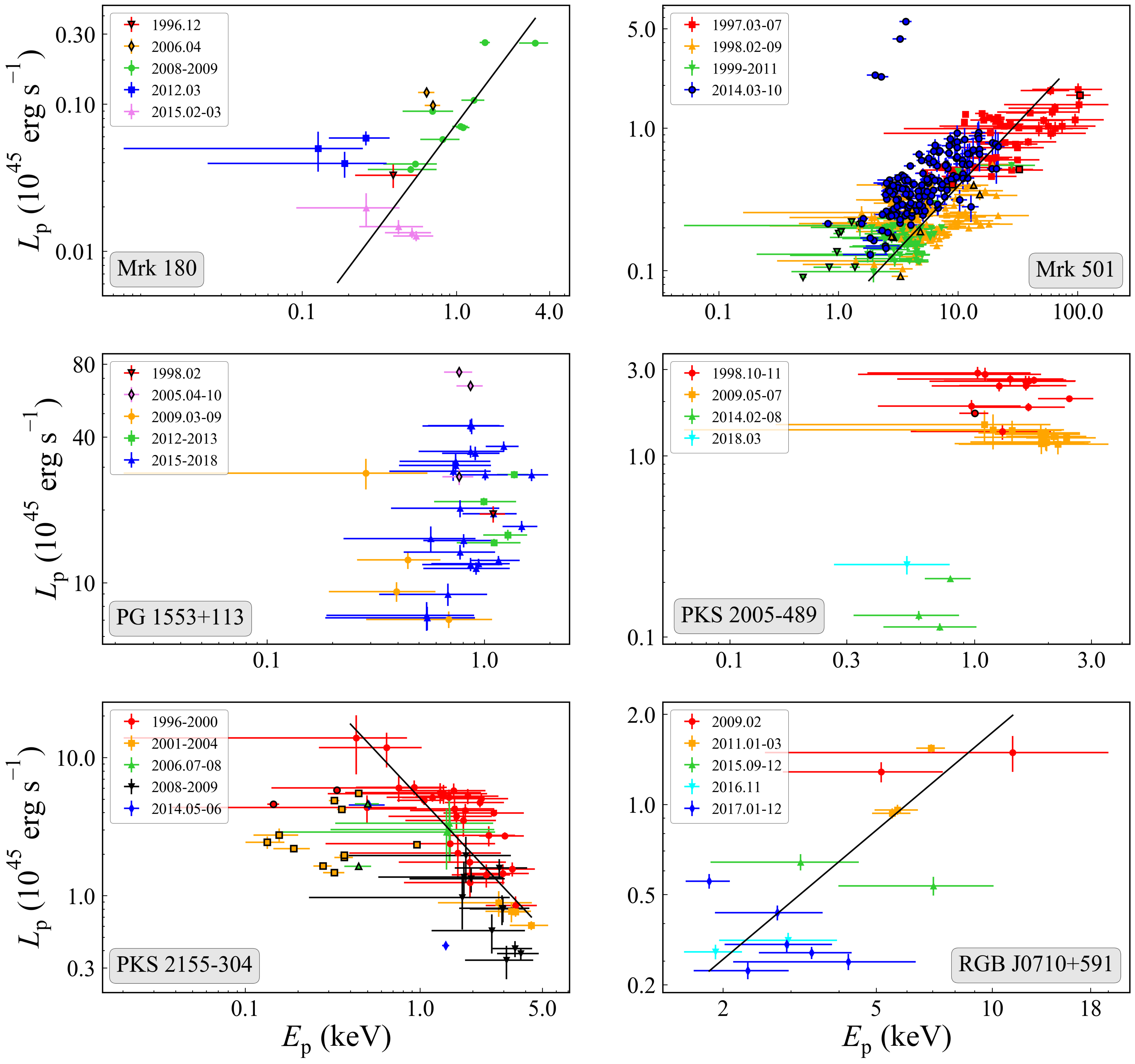}
\caption{}
\end{figure*}

\renewcommand{\thefigure}{\arabic{figure}}

\section{METHOD}
\label{sec:method}

We used the log-parabolic model \citep[][]{2004A&A...413..489M} to fit each \rxte\ or \swift\ spectrum with the {\sc{xspec}} software package (version 12.9.0). The log-parabolic model can describe curved spectra without invoking a sharp high-energy cut-off. \cite{2004A&A...413..489M} provided a physical explanation of this model in the framework of statistical acceleration process. This model has two forms in {\sc{xspec}}: {\tt{logpar}} and {\tt{eplogpar}}, both of which are used in this work. The Galactic hydrogen column density ($\textit{N}_\textrm{H}$) derived from \cite{Dickey_1990} for each source was fixed during the fitting process.

\subsection{{\tt{logpar}} model}
\label{sec:logpar}
The log-parabolic model is given by 
\begin{equation}
\textit{F(E)}=\textit{K}(\textit{E}/\textit{E}_1)^{(-\textit{a}-\textit{b} \log(\textit{E}/\textit{E}_1))},
\label{eq:logpar}
\end{equation} 
in units of $\textrm{photons}\ \textrm{cm}^{-\textrm{2}}\ \textrm{s}^{-\textrm{1}}\ \textrm{keV}^{-\textrm{1}}$ \citep[e.g.,][]{2004A&A...413..489M}. $\emph{E}_\textrm{1}$ is the reference energy, generally fixed to 1 keV. The parameter \aaa\ is the spectral index at the energy of $\emph{E}_\textrm{1}$, while \textit{b} is the curvature parameter around the peak. If \textit{b} = 0, it becomes a power-law model. \textit{K} is the normalization.
The location of the synchrotron peak is calculated by 
\begin{equation}
\textit{E}_{\textrm{p,logpar}}=\textit{E}_\textrm{1} \textrm{10}^{(\textrm{2}-\textit{a})/\textrm{2}\textit{b}}\ \ \ (\textrm{keV})
\label{eq:logep} 
\end{equation}
and the peak height is calculated by
\begin{align} 
\textit{S}_{\textrm{p,logpar}} &= \bigl(\textrm{1.60} \times \textrm{10}^{-\textrm{9}} \bigr)\ \textit{K} \textit{E}_\textrm{1} \textit{E}_{\textrm{p}} \bigl(\textit{E}_{\textrm{p}}/\textit{E}_\textrm{1} \bigr)^{-\textit{a}/\textrm{2}},   \\
                              &= \bigl(\textrm{1.60} \times \textrm{10}^{-\textrm{9}} \bigr)\ \textit{K} \textit{E}_\textrm{1}^\textrm{2}\ \textrm{10}^{(\textrm{2}-\textit{a})^\textrm{2}/\textrm{4}\textit{b}}\ \ \ \bigl(\textrm{erg}\ \textrm{cm}^{-\textrm{2}}\ \textrm{s}^{-\textrm{1}} \bigr).
\label{eq:logfp}
\end{align}

\subsection{{\tt{eplogpar}} model}
\label{sec:eplogpar}
Even though the maximum-likelihood estimates for \ep\ and $\textit{S}_\textrm{p}$ can be obtained using Eq. \ref{eq:logep} and Eq. \ref{eq:logfp}, their error propagation is complex. We thus adopted another form of the log-parabolic model, which is given by
\begin{equation}
\textit{F(E)}=\textit{K} \textrm{10}^{-\textit{b} (\log{(\textit{E}/\textit{E}_{\textrm{p}})})^\textrm{2}}/\textit{E}^\textrm{2},
\label{eq:eplogpar}
\end{equation}
in units of $\textrm{photons}\ \textrm{cm}^{-\textrm{2}}\ \textrm{s}^{-\textrm{1}}\ \textrm{keV}^{-\textrm{1}}$ \citep[e.g.,][]{2007A&A...466..521T,2009A&A...501..879T}. $\textit{E}_\textrm{p}$ is the synchrotron peak in units of keV (hereafter $\textit{E}_\textrm{p,eplog}$), while \textit{b} is the curvature parameter, which is the same as the parameter \textit{b} in the {\tt{logpar}} model. The parameter \textit{K} is the flux in $\nu F_\nu$ units at energy \ep\ keV. The synchrotron peak height is calculated by
\begin{equation}
\textit{S}_{\textrm{p,eplog}}=\textrm{1.60} \times \textrm{10}^{-\textrm{9}} \times \textit{K}\ \ \ \bigl(\textrm{erg}\ \textrm{cm}^{-\textrm{2}}\ \textrm{s}^{-\textrm{1}} \bigr).
\label{eq:eplogfp}
\end{equation}

\subsection{Spectral Analysis}
\label{sec:spec_ana}
Firstly, we fitted the spectra with the power-law and the {\tt{logpar}} models, and then used the F-test to compare the fitting results of these two models. 
In order to sift out the spectra preferring the log-parabolic model, we excluded the spectra with $p$-value larger than 0.05, because if a spectrum is well described by the power-law model, then the peak energy and curvature parameter could not be well constrained.
Secondly, we derived the peak parameters from the best-fit spectral parameters with the {\tt{logpar}} model using Eq. \ref{eq:logep} and Eq. \ref{eq:logfp}. 
Thirdly, we fitted the spectra with the {\tt{eplogpar}} model by setting the initial parameter values to those obtained from the previous step. Following the method in \cite{2009A&A...501..879T}, we used two criteria to test the reliability of the fitting results:
\begin{itemize}
 \item The ratio between $\textit{E}_\textrm{p,eplog}$ and 1-sigma uncertainty of $\textit{E}_\textrm{p,eplog}$ should be larger than 1 to assure the statistical significance of the peak energy.
 \item $\textit{E}_{\textrm{p,logpar}}$ should be consistent with $\textit{E}_\textrm{p,eplog}$ within 1-sigma uncertainty.
\end{itemize}
We excluded observations with $\textit{E}_\textrm{p,eplog}$ not satisfying the above criteria. 
Finally, we required that the parameter \textit{b} should be larger than 0, because when \textit{b} < 0, the resulting peak energy might straddle the cavity location of the concave spectrum, which might be the intersection of the two components from the synchrotron and inverse Compton scattering radiation processes, respectively.
The resulting numbers of \rxte\ and \swift\ observations adopted for each source in the total sample are shown in Table \ref{table:table1}, and these observations are annotated in the corresponding source light curves as presented in Figure~\ref{fig:lc}.

The rest-frame peak energy is calculated by
\begin{equation}
\textit{E}_{\textrm{p}}=(1+\emph{z})\textit{E}_{\textrm{p,eplog}}\ \ \ (\textrm{keV}).
\label{eq:peakenergy}
\end{equation}
The rest-frame isotropic peak luminosity is calculated by
\begin{equation}
\textit{L}_{\textrm{p}}\simeq \textrm{4}\pi \textit{D}_{\textrm{L}}^\textrm{2} \times \textit{S}_{\textrm{p,eplog}}\ \ \ \bigl(\textrm{erg}\ \textrm{s}^{-\textrm{1}} \bigr),
\label{eq:peaklumi}
\end{equation}
where $\textit{D}_\textrm{L}$ is the luminosity distance.

\section{RESULTS AND DISCUSSIONS}
\label{sec:results}

\begin{figure*}[t]
\centering
\includegraphics[width=\linewidth, clip]{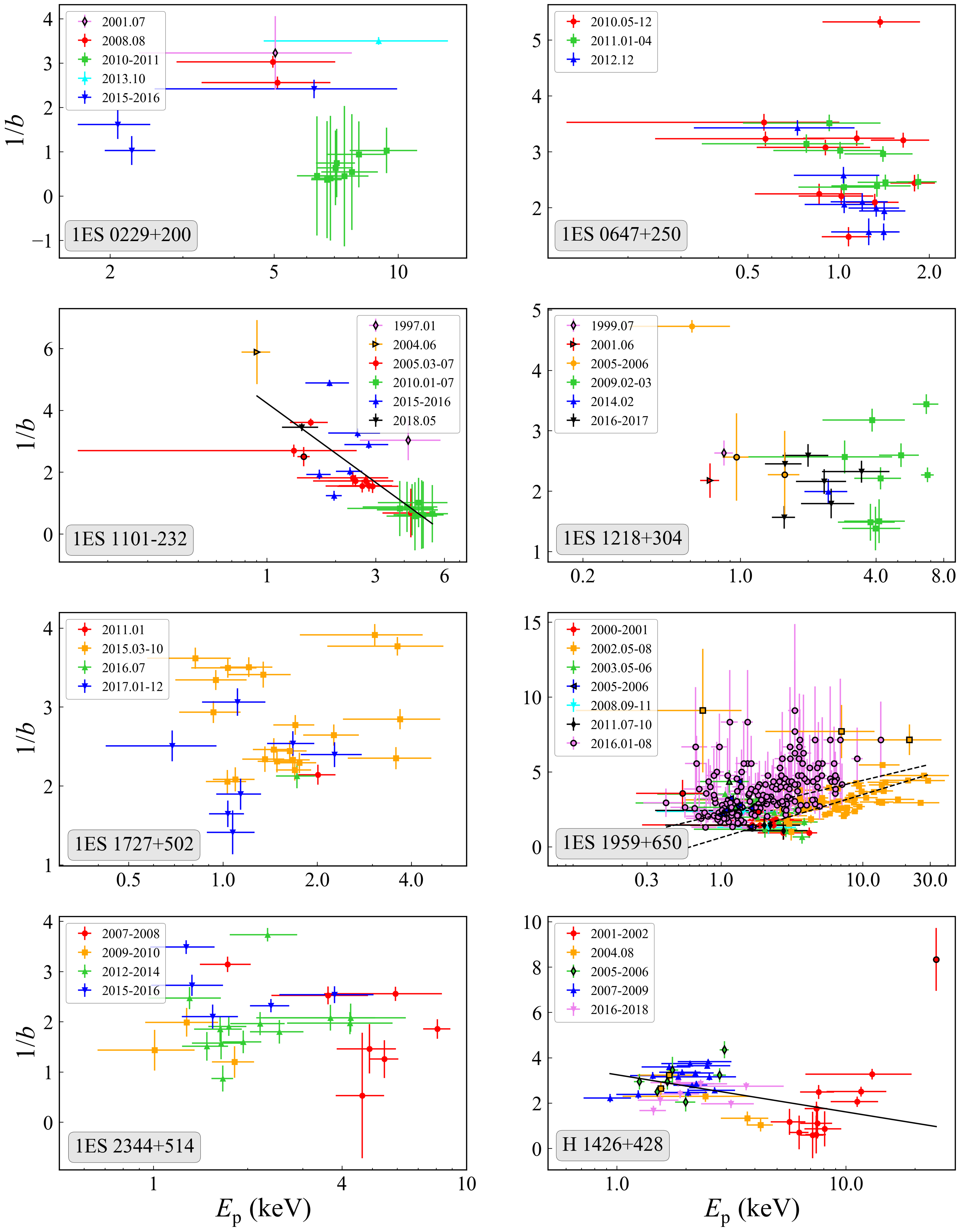}
\caption{Same as Figure \ref{fig:eplp}, but for \epb\ relations.}
\label{fig:epb}
\end{figure*}

We tested the correlations between \ep, \lp, and \bb\ in each source of the total sample with the Spearman's rank correlation coefficients (\rs) and associated $p$-values (\ps). 
For the objects displaying significant correlations (i.e., with \ps $\le$ 0.05), we used $\log {\textit{L}_\textrm{p}}=\alpha \log {\textit{E}_\textrm{p}}+\beta$ and $\textrm{1}/\textit{b}=\textit{C} \log {\textit{E}_\textrm{p}}+\textit{D}$ to describe the \eplp\ relation (see Figure \ref{fig:eplp}) and \epb\ relation (see Figure \ref{fig:epb}) following the fitting method of \cite{2007ApJ...665.1489K}, respectively. All the fitting results are shown in Table \ref{table:table2}.

\renewcommand{\thefigure}{\arabic{figure} (Cont.)}
\addtocounter{figure}{-1}

\begin{figure*}[t]
\centering
\includegraphics[width=\linewidth, clip]{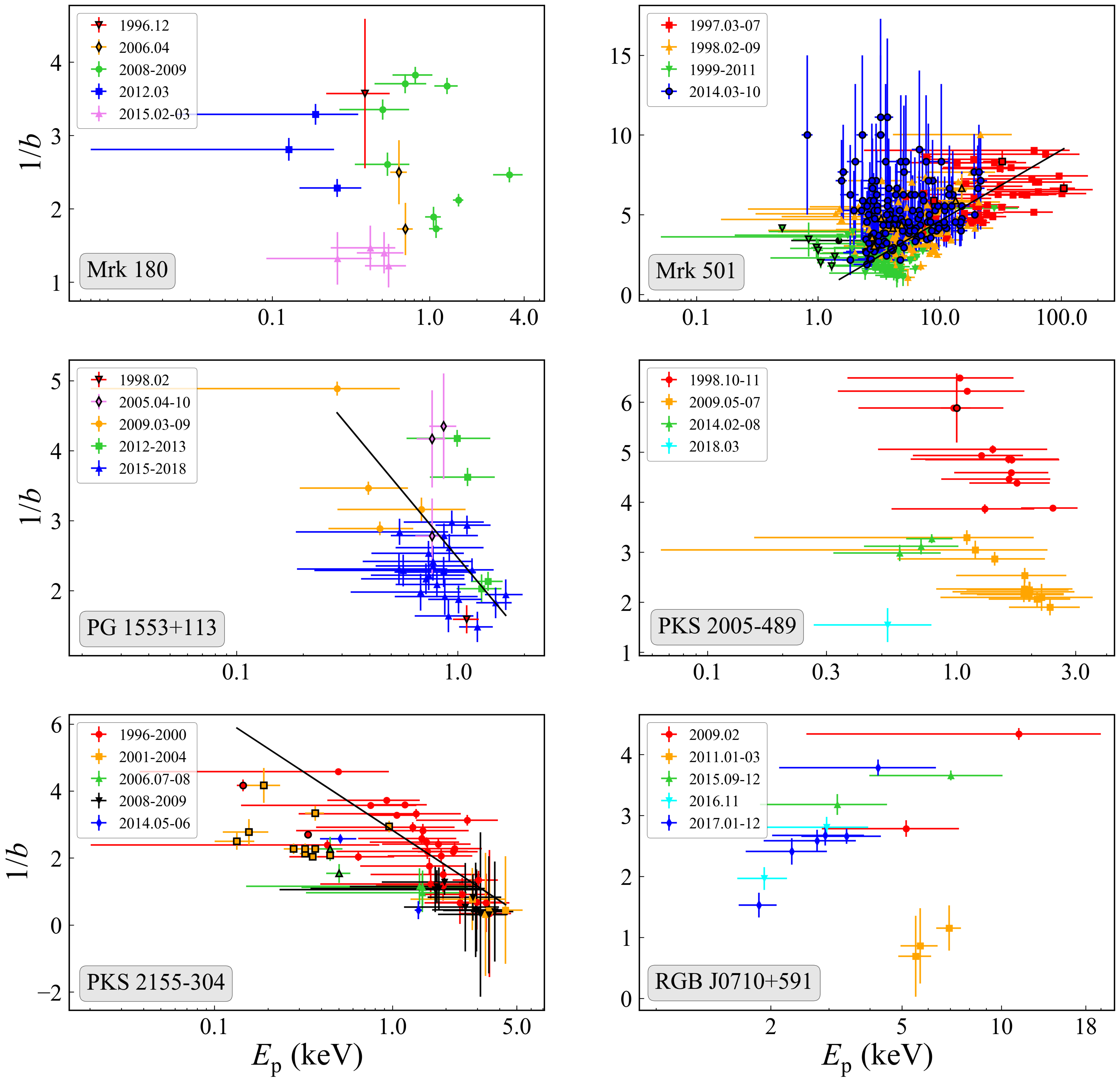}
\caption{}
\end{figure*}

\renewcommand{\thefigure}{\arabic{figure}}

\subsection{Properties of Individual Sources}
\label{sec:sources}

(1) 1ES~0229+200 had outbursts from 2010 to 2011, during which its peak luminosity reached $\textrm{10}^\textrm{45.1} \ \textrm{erg}\ \textrm{s}^{-1}$, peak energy reached 9.4 keV, and curvature maintained nearly at 1. 
The five observations in 2008, 2015 and 2016 represented the low-flux states of this source, whose peak luminosity decreased to $\textrm{10}^\textrm{44.3} \ \textrm{erg}\ \textrm{s}^{-1}$, peak energy decreased to 2 keV, and curvature decreased to about 0.5. Our result is consistent with that of \massaro.

(2) 1ES~0647+250 showed a gradually increasing trend of X-ray flux from 2010 to 2012. During this period, its peak luminosity increased from $\textrm{10}^\textrm{45.8}$ to $\textrm{10}^\textrm{46.5} \ \textrm{erg}\ \textrm{s}^{-1}$, while its peak energy ranged between 0.5 and 2 keV, and curvature around the peak ranged between 0.25 and 1. Given the large error bars of the fitting result of \eplp\ relation (see Table~\ref{table:table1}), we do not show the best-fit linear model of this source in Figure \ref{fig:eplp}.

(3) 1ES~1101-232 had an average peak energy of 2.5 keV, an average peak luminosity of $\textrm{10}^\textrm{45.6}\ \textrm{erg}\ \textrm{s}^\textrm{-1}$, and a curvature value of 0.5 in 2005. 
In 2010, its peak energy increased to 4.5 keV, peak luminosity decreased to $\textrm{10}^\textrm{45.3}\ \textrm{erg}\ \textrm{s}^\textrm{-1}$, and curvature increased to about 1. Between 2015 and 2016, its peak energy decreased to about 2 keV, peak luminosity increased to $\textrm{10}^\textrm{45.4}\ \textrm{erg}\ \textrm{s}^\textrm{-1}$, and the curvature decreased to 0.3 (similar to that in 2005). 
Our result based on the 2005 \rxte\ data is consistent with that of \massaro\ based on the 2005 \swift\ data.

(4) 1ES~1218+304 showed its peak luminosity increasing to $\textrm{10}^\textrm{46}$ $\textrm{erg}\ \textrm{s}^\textrm{-1}$ and peak energy increasing to 7 keV during an outburst in 2009. In 2005, 2016 and 2017, it was in relatively low-flux states, whose peak luminosity decreased to $\textrm{10}^\textrm{44.9}$ $\textrm{erg}\ \textrm{s}^\textrm{-1}$ and peak energy decreased to 0.6 keV. From 2005 to 2017, the curvature ranged between 0.25 and 1. 
It had a similar peak luminosity in 2005--2006 and 2016--2017, but had a lower peak energy range during the former period.

(5) 1ES~1727+502 showed its peak luminosity increasing from $\textrm{10}^\textrm{44}$ to $\textrm{10}^\textrm{44.5}$ $\textrm{erg}\ \textrm{s}^\textrm{-1}$ or even higher, and peak energy increasing from 0.8 to 3.5 keV during a large outburst in 2015. In the relatively low-flux state, its peak luminosity decreased to $\textrm{10}^\textrm{43.5}$ $\textrm{erg}\ \textrm{s}^\textrm{-1}$. From 2015 to 2017, its curvature had a very weak increase. 
It had a similar peak energy range in 2015 and 2017, but had a larger peak luminosity in 2015 than that in 2017.

(6) 1ES~2344+514 underwent a large outburst in December 2007, and the observation with the highest peak luminosity and peak energy in the \eplp\ plot corresponds to the peak position of this flare. During this period, its peak luminosity reached $\textrm{10}^\textrm{44.6}$ $\textrm{erg}\ \textrm{s}^\textrm{-1}$ and peak energy reached 8 keV. The three observations between 2009 and 2010 represented the low-flux states, whose peak luminosity decreased to $\textrm{10}^\textrm{43.3}$ $\textrm{erg}\ \textrm{s}^\textrm{-1}$ and peak energy decreased to 1 keV. In addition, its curvature displayed no significant changes.

(7) 1ES 1959+650 was in large outbursts in 2002 and 2016. In 2002 (\rxte\ observations), its peak luminosity increased from $\textrm{10}^\textrm{44.3}$ to $\textrm{10}^\textrm{45.5}$ $\textrm{erg}\ \textrm{s}^\textrm{-1}$ and peak energy increased from 0.7 to 30 keV. In 2016 \citep[\swift\ observations of][]{2018MNRAS.473.2542K}, its peak luminosity increased from $\textrm{10}^\textrm{44.5}$ to $\textrm{10}^\textrm{45.3}$ $\textrm{erg}\ \textrm{s}^\textrm{-1}$ and peak energy increased from 0.4 to 13 keV. In the same peak energy range, the peak luminosity in 2016 was five times larger than that in 2002. 
For the observations in both 2002 and 2016, there is a significant positive correlation between \ep\ and \lp\ as well as between \ep\ and \bb. The observations in 2005 and 2006 showed a similar trend with that in 2016. 
Interestingly, in 2000, 2001, 2003, 2008 and 2011, \ep\ and \lp\ seem to stay in a transitional region between the relations in 2002 and 2016 in the \eplp\ plot (see Figure \ref{fig:eplp}).

(8) H~1426+428 went through a large outburst in May 2001, which was excluded by one of our observation selection criteria (i.e., the {\tt{logpar}} model was not required to fit the spectra as the synchrotron peak was much larger than the \rxte\ spectral coverage; see Section~\ref{sec:spec_ana}); during the peak of this flare, its peak energy and peak luminosity were estimated to be larger than 25 keV and $\textrm{10}^\textrm{45.6}$ $\textrm{erg}\ \textrm{s}^\textrm{-1}$, respectively. 
In 2002, its peak energy ranged between 5 and 25 keV, and peak luminosity ranged between $\textrm{10}^\textrm{44.6}$ and $\textrm{10}^\textrm{45.3}$ $\textrm{erg}\ \textrm{s}^\textrm{-1}$. After 2004, its peak energy ranged between 0.9 and 5 keV, which is lower than that in 2002, but its peak luminosity ranged between $\textrm{10}^\textrm{44.7}$ and $\textrm{10}^\textrm{45.3}$ $\textrm{erg}\ \textrm{s}^\textrm{-1}$, which is quite similar to that in 2002. Our result is consistent with that of \massaro.

(9) Mrk~180 was in a high-flux state between 2008 and 2009. During this period, its peak energy changed between 0.5 and 3 keV, while peak luminosity changed between $\textrm{10}^\textrm{43.5}$ and $\textrm{10}^\textrm{44.5}$ $\textrm{erg}\ \textrm{s}^\textrm{-1}$. In 2015, it was in a relatively low-flux state with peak luminosity decreasing to $\textrm{10}^\textrm{43}$ $\textrm{erg}\ \textrm{s}^\textrm{-1}$. Our result is consistent with that of \massaro.

(10) Mrk~501 experienced a large outburst in 1997, whose peak luminosity reached $\textrm{10}^\textrm{45.3}$ $\textrm{erg}\ \textrm{s}^\textrm{-1}$ and peak energy could reach 100 keV. From 1997 to 2011, its peak energy decreased from 100 to 0.4 keV, while peak luminosity decreased from $\textrm{10}^\textrm{45.3}$ to $\textrm{10}^\textrm{44}$ $\textrm{erg}\ \textrm{s}^\textrm{-1}$. 
Our result based on the \rxte\ data between 1997 and 2011 is consistent with that of \massaro\ based on the \beppo\ and \swift\ data.
In addition, our result is also consistent with that of \cite{2017MNRAS.469.1655K}.

(11) PG~1553+113 showed its peak luminosity ranging between $\textrm{10}^\textrm{45.8}$ and $\textrm{10}^\textrm{46.6}$ $\textrm{erg}\ \textrm{s}^\textrm{-1}$, and peak energy ranging between 0.5 and 1.5 keV from 2015 to 2018. Our result is similar to that of \massaro.

(12) PKS~2005-489 was in a large outburst in 1998, whose peak luminosity reached $\textrm{10}^\textrm{45.3}$ $\textrm{erg}\ \textrm{s}^\textrm{-1}$. In 2014, it was in a relatively low-flux state with peak luminosity decreasing to $\textrm{10}^\textrm{44}$ $\textrm{erg}\ \textrm{s}^\textrm{-1}$ and the curvature value similar to that of the high-flux state in 2009. Its peak energy ranged between 0.5 and 2.5 keV. 
Our result based on the 1998 \rxte\ data is consistent with that of \massaro\ based on the 1998 \beppo\ data.
Given the large error bars of the fitting result of \epb\ relation, we do not show the best-fit linear model in Figure \ref{fig:epb}.

(13) PKS 2155-304 underwent at least two large outbursts between 1996 and 2000, whose peak luminosity reached $\textrm{10}^\textrm{46}$ $\textrm{erg}\ \textrm{s}^\textrm{-1}$ and peak energy ranged between 0.5 and 3 keV. The observations in 2009 were in a relatively low-flux state, whose peak luminosity decreased to $\textrm{10}^\textrm{44.5}$ $\textrm{erg}\ \textrm{s}^\textrm{-1}$ but peak energy reached 4 keV. Within uncertainties, our result is consistent with that of \massaro.

(14) RGB J0710+591 showed its peak luminosity reaching $\textrm{10}^\textrm{45.2}$ $\textrm{erg}\ \textrm{s}^\textrm{-1}$ and peak energy reaching 10 keV in the outbursts of 2009 and 2011, while the curvature in 2011 was much higher than that in 2009. In the low-flux state in 2017, its peak luminosity decreased to $\textrm{10}^\textrm{44.5}$ $\textrm{erg}\ \textrm{s}^\textrm{-1}$ and peak energy decreased to 2 keV.

\subsection{$\textit{E}_\textrm{p}$-$\textit{L}_\textrm{p}$ Relation}
\label{sec:eplp}

According to the synchrotron theory \citep{1979AstQ....3..199R}, the synchrotron peak energy ($\textit{E}_\textrm{p}$) and luminosity ($\textit{L}_\textrm{p}$) follow a power-law relation of $\textit{L}_\textrm{p} \propto \textit{E}_\textrm{p}^{\alpha}$. If the electrons in the emitting region follow a log-parabolic distribution, the peak luminosity is given by $\textit{L}_\textrm{p} \propto \emph{n}\textrm{(}\gamma_\textrm{3p}\textrm{)}\gamma_\textrm{3p}^\textrm{3} \textit{B}^\textrm{2} \delta^\textrm{4} \sim \textit{N} \gamma_{\textrm{p}}^\textrm{2} \textit{B}^\textrm{2} \delta^\textrm{4}$, and the peak energy follows $\textit{E}_\textrm{p} \propto \gamma_{\textrm{3p}}^\textrm{2} \textit{B} \delta \sim \gamma_{\textrm{p}}^\textrm{2} \textit{B} \delta$ \citep[e.g.,][]{2009A&A...501..879T}. $\gamma_{\textrm{3p}}$ represents the peak of $\textit{n}\textrm{(}\gamma\textrm{)} \gamma^3$ where $\gamma$ is the electron Lorentz factor, $\gamma_{\textrm{p}}$ represents the peak of $\textit{n}\textrm{(}\gamma\textrm{)} \gamma$, $\textit{N} \sim \emph{n}(\gamma_\textrm{p})\gamma_\textrm{p}$ is the total electron number, \textit{B} is the magnetic field and $\delta$ represents the Doppler beaming factor. 

If $\alpha$ = 1, the spectral changes might be mainly caused by the variations of the average electron energy, and the total electron number remains constant; if $\alpha$ = 1.5, the spectral changes might be mainly caused by the variations of the average electron energy, but the total electron number also changes; if $\alpha$ = 2, the spectral variations might be correlated with the changes of the magnetic field; and if $\alpha$ = 4, the spectral changes might be then dominated by the variations of the beaming factor.

According to the fitting results (see Table \ref{table:table2}), five objects (i.e., 1ES~0229+200, 1ES~1727+502, 1ES~1959+650 in 2002, 1ES~2344+514, RGB J0710+591) have $\alpha\approx 1$, thus their spectral variations are mainly caused by the changes of the electron energy while the total electron number may remain constant. 
The spectral variations of 1ES~1218+304, 1ES~1727+502 and Mrk~180 might also be due to the changes of the electron energy but the total electron number changes. 
1ES~1959+650 in 2016 and Mrk~501 have $\alpha<1$, which could not be explained by any mechanism mentioned above. 
In addition, both 1ES~1101$-$232 and PKS~2155$-$304 show an anti-correlation between \ep\ and \lp, which is significantly different from the other sources. H 1426+428, PG~1553+113 and PKS 2005$-$489 show no correlation between \ep\ and \lp. For 1ES~0647+250, due to the large errors of $\alpha$, we could not draw any solid conclusion.

In a word, most of the sources (9/14) show a positive correlation between \ep\ and \lp, which indicates that their spectral variations might be due to the variations of electron energies. For the sources that show a negative or no correlation between \ep\ and \lp, the aforementioned mechanisms could not explain their results and it might be related to the source luminosity (see Section \ref{totalsample}). In addition, 1ES~1959+650 has two different \eplp\ relations in 2002 and 2016, indicating that the causes of spectral variations likely changed during these two periods.

\begin{figure*}[!tp]
\centering
\includegraphics[width=0.9\linewidth, clip]{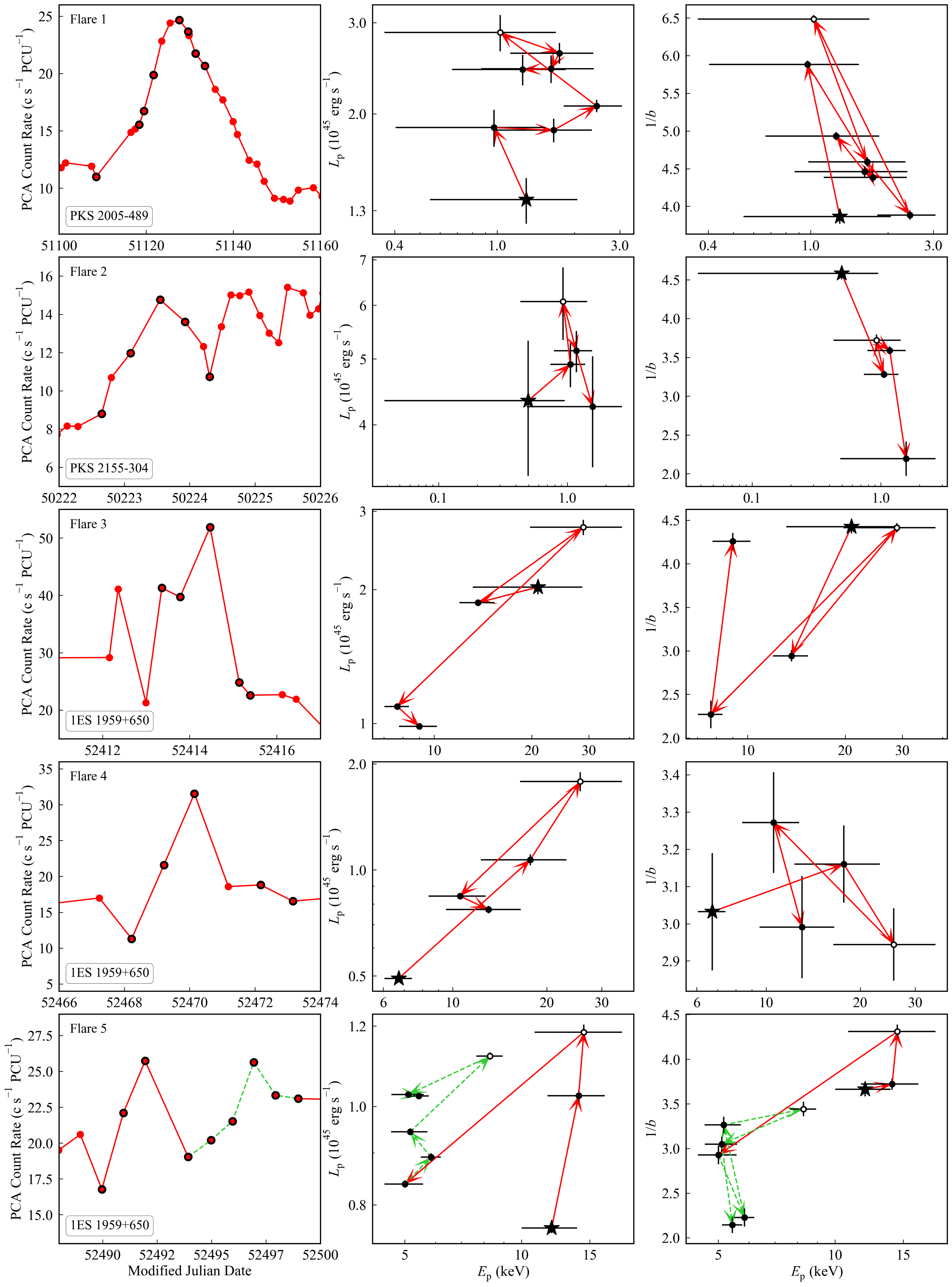}
\caption{3--25 keV \rxte/PCA light curve (first column), time evolution of \eplp\ and \epb\ relations (second and third columns, respectively) for PKS~2005$-$489 (Flare 1), PKS~2155$-$304 (Flare 2), 1ES~1959+650 (Flares 3--5) and Mrk~501 (Flares 6--10), respectively. 
In the first column, the filled circles with black edges of the light curves represent the observations satisfying the selection criteria in Sections \ref{sec:samcon} and \ref{sec:spec_ana}. 
In the second and third columns, the beginning of each flare is marked with a star, the peak of each flare is marked with an open circle, the red and green arrowed lines represent the first and second flares (see the first column) respectively, and the arrows indicate the time order. It seems that the \eplp\ relation does not exhibit significant differences between flares, while the \epb\ relation differs from flare to flare.}
\label{fig:flare_eplpb}
\end{figure*}

\renewcommand{\thefigure}{\arabic{figure} (Cont.)}
\addtocounter{figure}{-1}

\begin{figure*}[h]
\centering
\includegraphics[width=0.9\linewidth, clip]{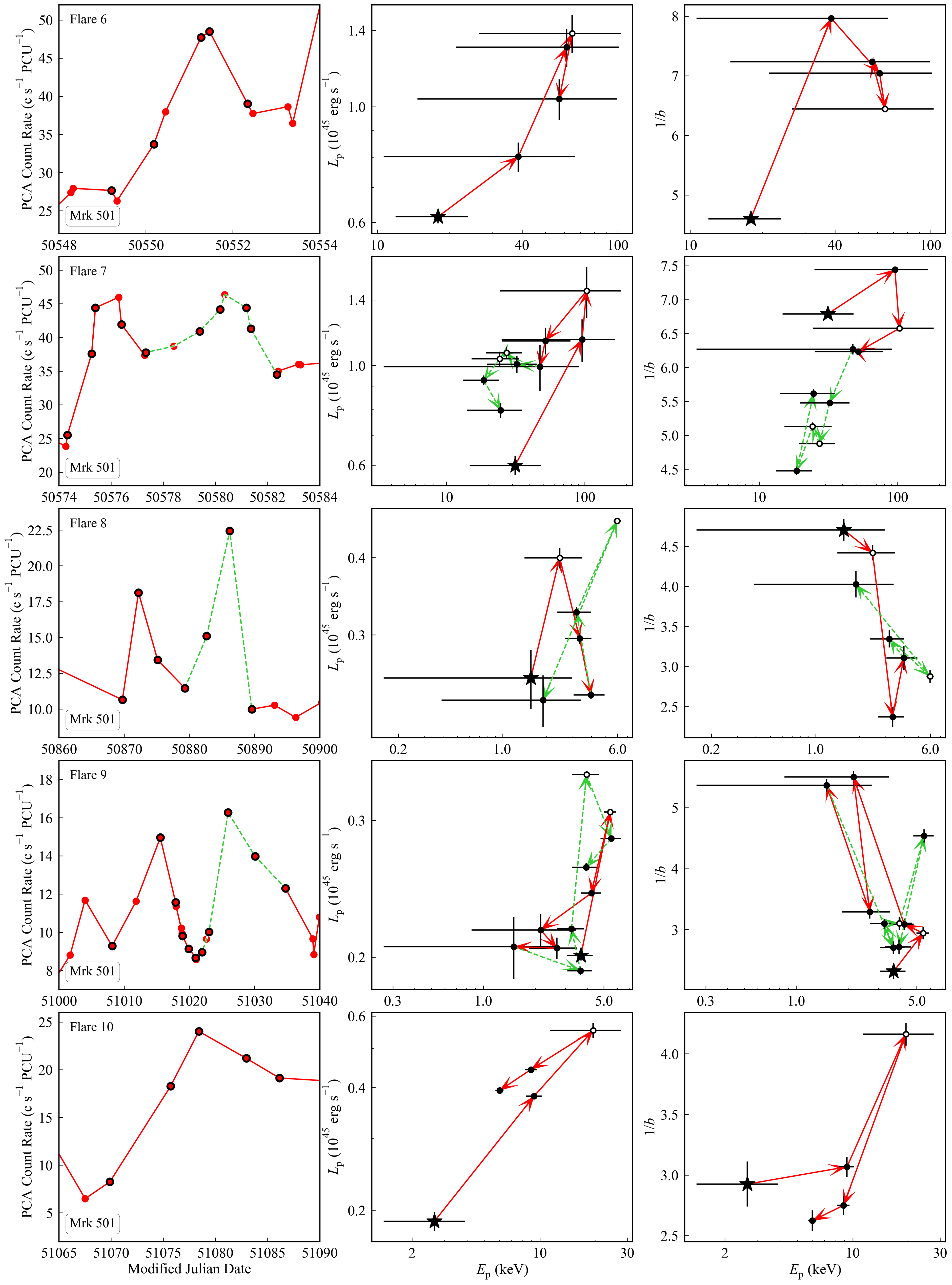}
\caption{}
\end{figure*}
\renewcommand{\thefigure}{\arabic{figure}}

\begin{figure*}[t]
\centering
\includegraphics[width=0.8\linewidth, clip]{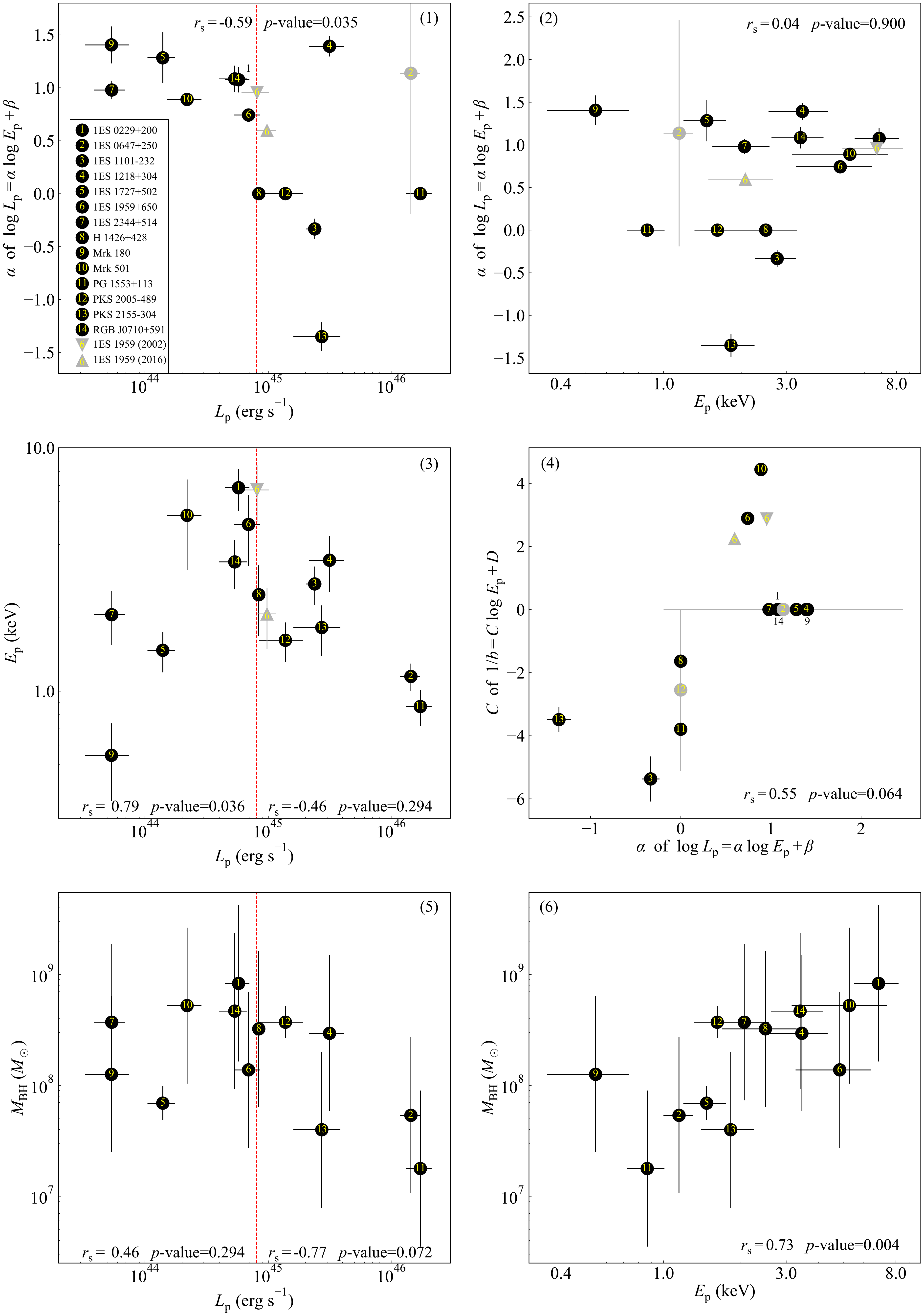}
\caption{Correlations between $\alpha$ and \lp\ (panel 1), between $\alpha$ and \ep\ (panel 2), between \ep\ and \lp\ (panel 3), between \textit{C} and $\alpha$ (panel 4), between $M_\textrm{BH}$ and \lp\ (panel 5), and between $M_\textrm{BH}$ and \ep\ (panel 6) for the 14 sources in the total sample. The black filled circles represent the median values of \ep\ and \lp\ of respective all observations as well as best-fitting \textit{C} and $\alpha$ of each source. 
The triangle and inverted-triangle symbols represent the data of 1ES~1959+650 in 2016 and 2002, respectively. The red dashed line annotates that \lp\ = $8 \times \textrm{10}^\textrm{44}\ \textrm{erg}\ \textrm{s}^\textrm{-1}$.
The error bars of \ep\ and \lp\ represent the standard deviations. 
In the panel (4), the Spearman's rank test is performed without Sources 2 and 12 due to their large error bars. 
In panels (1) and (2), the Spearman's rank test is performed without Source 2 due to its large error bars. 
}
\label{fig:totalsource}
\end{figure*}

\subsection{$\textit{E}_\textrm{p}$-$\textit{b}$ Relation}
\label{sec:epandb}

It has been suggested that the statistical and stochastic acceleration mechanisms could explain the correlation between $\textit{E}_\textrm{p}$ and \textit{b} \citep[e.g.,][]{2004A&A...413..489M,2011ApJ...739...66T}. These two mechanisms can produce the electron energy distribution that follows the log-parabolic law, resulting in a log-parabolic SED.

The first scenario is described by the statistical acceleration process. For the energy-dependent acceleration probability process, the electron energy distribution follows the log-parabolic law and the acceleration efficiency of the particles is inversely proportional to their energy \citep{2004A&A...413..489M}. In this process, $\textit{E}_\textrm{p}$ and \textit{b} follow the correlation of $\log \textit{E}_\textrm{p} \approx \textit{Const.}+\textrm{2}/\textrm{(5}\textit{b}\textrm{)}$, given the assumption of \textit{b} = \textit{r}/4 where $r$ is the curvature of the electron energy distribution \citep[][]{2014ApJ...788..179C}. 
While for the fluctuations of fractional acceleration gain process, electron energies are distributed in a log-normal law, and the energy gain fluctuations are a random variable around the systematic energy gain \citep{2011ApJ...739...66T}. In this case, $\textit{E}_\textrm{p}$ and \textit{b} follow the correlation of $\log \textit{E}_\textrm{p} \approx \textit{Const.}+\textrm{3}/\textrm{(10}\textit{b}\textrm{)}$ given \textit{b} = \textit{r}/4 \citep[][]{2014ApJ...788..179C}. 

Another scenario is described by the stochastic acceleration process, and its kinetic equation includes a momentum-diffusion term, which leads to energy gain fluctuations in the diffusive shock acceleration process \citep{2011ApJ...739...66T}. For this explanation, $\textit{E}_\textrm{p}$ and \textit{b} follow the relation of $\log \textit{E}_\textrm{p} \approx \textit{Const.}+\textrm{1}/\textrm{(2}\textit{b}\textrm{)}$ given \textit{b} = \textit{r}/4 \citep[][]{2014ApJ...788..179C}.

Therefore, the theoretically expected values of \bigc\ are 10$/$3, 5$/$2 and 2 for the fractional acceleration gain fluctuation, energy-dependent acceleration probability and stochastic acceleration processes, respectively.

However, according to the fitting results (see Table \ref{table:table2}), only Mrk~501 and 1ES~1959+650 show a significant positive correlation between \ep\ and \bb. For 1ES~1959+650 in 2002, \bigc$=$2.87$\pm$0.23, which is closely consistent with the energy-dependent acceleration probability scenario; for 1ES~1959+650 in 2016, \bigc$=$2.25$\pm$0.14, which indicates that either the energy-dependent acceleration probability scenario or the stochastic acceleration process could be at work. For Mrk~501, \bigc$=$4.44$\pm$0.15, which can not be explained by any aforementioned mechanism. In addition, the following five sources show an anti-correlation between \ep\ and \bb: 1ES~1101--232, H~1426+428, PG~1553+113, PKS~2005$-$489 and PKS~2155$-$304. In contrast, there is no correlation between \ep\ and \bb\ in seven sources, i.e., 1ES~0229+200, 1ES~0647+250, 1ES~1218+304, 1ES~1727+502, 1ES~2344+514, Mrk~180 and RGB~J0710+591. For PKS~2005$-$489, due to the large errors of $C$, we could not draw any solid conclusion. 

None of these three mechanisms could explain all the \epb\ behaviors of these sources. One possible reason is that the \epb\ relation might be different from flare to flare in each source (see Section \ref{flares}), as different acceleration processes might be at work during flares and hence cause large scatters of the \epb\ relation for each source. Therefore, we would not expect a simple and good \epb\ relation that can be explained by one single acceleration model.

\subsection{Correlations during the Single Flares}
\label{flares}

\cite{2004ApJ...601..759T} found that for Mrk~421, \ep\ and \lp\ showed an overall positive correlation but this relation seemed to vary between individual flares lasting for hundreds of kilo-seconds. We here focus on the flares lasting for several days. We require that in each flare, at least four observations should satisfy the selection criteria in Sections \ref{sec:samcon} and \ref{sec:spec_ana}. Finally, we selected out ten flares of four objects that were observed by \rxte\ (see Figure \ref{fig:flare_eplpb}).

For Flares 3--5 of 1ES~1959+650 and Flares 6--10 of Mrk~501 in Figure \ref{fig:flare_eplpb}, there are positive correlations between \ep\ and \lp, which are consistent with the trends revealed with their respective all observations (see Figure \ref{fig:eplp}). 
In addition, most of these flares share a similar slope of \eplp\ relation to that for their respective all observations, which indicates that the \eplp\ relation did not vary significantly during these individual flares. 
During these flares, the peak energy shifted to the higher energy and reached the highest value at the peak of flare, then returned to the lower energy with the flux decreasing. 
Assuming that the flaring events are due to the contribution from a new component, the positive correlation between \ep\ and \lp\ might indicate that this new component has a higher synchrotron peak energy than the preexisting component. 
In addition, in some cases, the observations during the rising periods had the higher peak energy compared with that in the decay periods, such as, Flares 4, 5, 7, 9, 10 and the second flare of Flare 8; while other cases show the opposite trend, such as, Flares 3, 6 and the first flare of Flare 8. For Flare 1 of PKS~2005$-$489 and Flare 2 of PKS~2155$-$304, \ep\ and \lp\ seem to follow the negative or no correlation, but given the large error bars, we could not draw any solid conclusion. For Flare 5 of 1ES~1959+650 and Flares 7--9 of Mrk~501, there are two adjacent and comparable flares, where the peak energies of the second flares of Flares 5 and 7 are lower than that of their respective first flares, while the first and second flares of Flares 8 and 9 have similar peak energies. 
Given that the two flares of Flare 5 or 7 lasted for about 2--4 days while the two flares of Flare 8 or 9 lasted for about 10--20 days, it is likely that the two flares of Flares 5 and 7 are from two different small-scale regions while the two flares of Flares 8 and 9 are from the same large-scale region.

For Flare 1 of PKS~2005$-$489, Flare 2 of PKS~2155$-$304, Flare 4 of 1ES~1959+650, and Flares 8 and 9 of Mrk~501, there is a negative correlation between \ep\ and \bb, while for Flares 3 and 5 of 1ES~1959+650 and Flares 6, 7, and 10 of Mrk~501, it shows an opposite trend. Therefore, the \epb\ relation differs from flare to flare for the same source. 
For Flares 1, 3, 5 and 10, the peak of the flare has the lowest curvature around the synchrotron peak compared with that of other observations during flares, while for other cases, the peak of the flare has a medium curvature value. The changes of curvature around the peak do not follow the flux variation trend during flares.

In conclusion, for the same source (Mrk~501 and 1ES~1959+650), the \eplp\ relation does not show any significant change between different flares lasting for several days, which is inconsistent with the result of shorter-timescale (i.e., hundreds of kilo-seconds) flares in \cite{2004ApJ...601..759T}. However, the \epb\ relation differs from flare to flare, which might explain the lack of correlation between \ep\ and \bb\ in most objects, or the large scatters of \epb\ relations.

\subsection{Correlations for the Total Sample}
\label{totalsample}

In Figure \ref{fig:totalsource}, we show the correlations between $\alpha$ and \lp, $\alpha$ and \ep, \ep\ and \lp, \textit{C} and $\alpha$, black hole mass (\bhmass) and \lp, and \bhmass\ and \ep\ of all the sources (using their respective median values of \ep\ and \lp\ as well as best-fitting \textit{C} and $\alpha$), respectively. Given that the small sample size of this work, we assume that the Doppler factors are similar for all the HBLs in our sample.

In the panel (1) of Figure \ref{fig:totalsource}, there seems to be an anti-correlation between \lp\ and $\alpha$, which indicates that the causes of spectral variations might be different between luminous and faint sources. For the sources with \lp\ lower than 8$\times \textrm{10}^{\textrm{44}}\ (\sim\textrm{10}^{\textrm{44.9}})\ \textrm{erg}\ \textrm{s}^{\textrm{-1}}$, they have the $\alpha$ value larger than 0.5, while for the sources with higher \lp, they usually have much smaller $\alpha$ values. Therefore, we divided the total sample into two subsamples according to the peak luminosity: low \lp\ sources (\lp\ < 8$\times \textrm{10}^{\textrm{44}}\ \textrm{erg}\ \textrm{s}^{\textrm{-1}}$, hereafter LLP) and high \lp\ sources (\lp\ > 8$\times \textrm{10}^{\textrm{44}}\ \textrm{erg}\ \textrm{s}^{\textrm{-1}}$, hereafter HLP). In addition, it seems that there is no correlation between \ep\ and $\alpha$ (see the panel 2 of Figure \ref{fig:totalsource}), which indicates that the causes of spectral variations show no significant difference between sources with different peak energies.

In the panel (3) of Figure \ref{fig:totalsource}, for HLP subsample, there seems to be an anti-correlation between \ep\ and \lp, but it is not statistically significant. While for LLP subsample, there is an apparent positive correlation between \ep\ and \lp. In addition, the trend of \eplp\ correlation in each individual object is consistent with that of the subsample it belongs to. For example, Mrk~180 shows a positive \eplp\ correlation ($\alpha$ > 0), and it belongs to LLP subsample, which also shows a positive \eplp\ correlation. These results might be explained by that with the number of high-energy electrons increasing, \ep\ will become larger and the number of photons produced by synchrotron radiation will increase (\lp\ will increase). However, when \lp\ is larger than 8$\times \textrm{10}^{\textrm{44}}\ \textrm{erg}\ \textrm{s}^{\textrm{-1}}$, the synchrotron self-Compton (SSC) cooling effect might be significant, which might result in a trend that \ep\ decreases with increasing \lp.

We test the correlation between the causes of the spectral variations and possible acceleration processes by studying the correlation between $\alpha$ and \textit{C} (see the panel 4 of Figure \ref{fig:totalsource}), because $\alpha$ and \textit{C} might represent the causes of the spectral variations and the acceleration processes, respectively. However, it seems that there is no significant correlation between $\alpha$ and \textit{C} according to the statistical test.

There is no significant correlation between \lp\ and \bhmass\ (from previous works, see Table \ref{table:table1}) either in the total sample or in the two subsamples, while \ep\ shows a significant positive correlation with \bhmass\ (see panels 5 and 6 of Figure \ref{fig:totalsource}). Our result is different from that of \cite{2015MNRAS.450.3568X}, which suggested that there is no apparent correlation between synchrotron peak frequency and black hole mass in a sample of more than 100 BL Lacs. There are several reasons that might cause such a discrepancy. Firstly, we mainly focus on HBLs, while they focus on all types of BL Lacs. Secondly, they fitted broadband SEDs or used empirical relationships to obtain \ep\ and \lp, while we fitted the X-ray spectra of each observation of each source, and then obtained median \ep\ and \lp\ values. In addition, the different methods of estimating \bhmass\ might also affect the result.

Finally, we also study the correlations between \ep\ and \bb, \lp\ and \bb, but it seems that there is no correlation. \cite{2014ApJ...788..179C} fitted the broadband SEDs of 48 blazars with two log-parabolic models and found a positive correlation between the synchrotron peak and 1/\textit{b} for the whole sample. \cite{2016MNRAS.463.3038X} used a larger sample including 200 FSRQs and 79 BL Lacs, and found a similar trend of \epb\ relation but the slope of this relation is different between FSRQs and BL Lacs. Our result seems to be inconsistent with theirs, which might be due to the following reasons: (1) both of their samples only included two or three sources with synchrotron peaks larger than 0.3 keV where our sources mainly peak; (2) their curvature values were obtained by fitting the broadband SEDs, while our result is obtained using the X-ray data, which usually shows larger curvature values. However, the result of \cite{2016MNRAS.463.3038X} showed that the \epb\ relation has a large scatter for HBLs, which might lead to a weak \epb\ correlation for the high-peaked sources.

\section{SUMMARY AND CONCLUSIONS}
\label{sec:summary}

For the 14 BL Lacs in the total sample (see Table \ref{table:table1}), we utilized the log-parabolic model to fit the 3--25 keV \textit{RXTE}/PCA and 0.3--10 keV \swift/XRT spectra, and then obtained the following three parameters to characterize the synchrotron peak: peak energy ($\textit{E}_\textrm{p}$), peak luminosity ($\textit{L}_\textrm{p}$) and the curvature parameter around the peak (\textit{b}). Further, we studied the \eplp\ and \epb\ relations in these sources and their trends during flaring periods. We also analyzed these correlations for the total sample and the correlations with the black hole mass.

The results regarding the \eplp\ relations of individual sources are as follows: (1) The positive \eplp\ relations of 1ES~0229+200, 1ES~1727+502, 1ES~1959+650 in 2002, 1ES~2344+514, and RGB~J0710+591 indicate that their spectral variations might be related to the electron energy variations, while the total electron number remains constant; and the positive \eplp\ relations of 1ES~1218+304, 1ES~1727+502, and Mrk~180 indicate that their spectral variations might be related to the electron energy variations but the total electron number changes. (2) 1ES~1959+650 in 2016 and Mrk~501 show a positive \eplp\ relation but none of the four aforementioned mechanisms could explain their spectral variations. (3) H~1426+428, PKS~2005$-$489 and PG~1553+113 show no correlation between \ep\ and \lp. (4) 1ES~1101$-$232 and PKS~2155$-$304 show an anti-correlation between \ep\ and \lp. None of the four mechanisms could explain the causes of spectral variations in these sources. (5) 1ES~1959+650 shows two significantly different \eplp\ relations in 2002 and 2016.

The results on the \epb\ relations of individual sources are the following: (1) Mrk~501 and 1ES~1959+650 show the positive correlation between \ep\ and \bb. For 1ES~1959+650, both the energy-dependent acceleration probability and stochastic acceleration processes could be the possible mechanisms. (2) The following five sources show an anti-correlation: 1ES~1101$-$232, H~1426+428, PG~1553+113, PKS~2005$-$489 and PKS~2155$-$304. (3) The following seven sources show no correlation: 1ES~0229+200, 1ES~0647+250, 1ES~1218+304, 1ES~1727+502, 1ES~2344+514, Mrk~180 and RGB~J0710+591. Except 1ES~1959+650, none of the aforementioned acceleration mechanisms could explain the correlations between \ep\ and \bb\ for the sources in the total sample.

For the total sample, $\alpha$ shows an anti-correlation with \lp, which indicates that the causes of spectral variations might be different between luminous and faint sources. In contrast, $\alpha$ shows no correlation with \ep. When \lp\ is lower than $8 \times \textrm{10}^{\textrm{44}}\ \textrm{erg}\ \textrm{s}^{\textrm{-1}}$, there is a significant positive correlation between \ep\ and \lp. In addition, \ep\ shows a positive correlation with \bhmass, while \lp\ shows no correlation with \bhmass.

During flares lasting for several days, the \eplp\ relation does not exhibit the significant change between different flares, which is not consistent with the previous result based on shorter-timescale (hundreds of kilo-seconds) flares. The \epb\ relation differs from flare to flare, which might explain the lack of correlation between \ep\ and \bb\ in most objects.

In the future and in light of our results in this work, we will use more data and larger samples when available to further investigate the correlations between the synchrotron peak parameters (\ep, \lp, and \bb) and the black hole properties (e.g., \bhmass) of blazars, aiming to better confront observational results with theoretical considerations and explore whether the correlation between the synchrotron \ep\ and \bhmass\ could provide a method to estimate the black hole mass of blazars.

\acknowledgements 
We are grateful to Zhenyi Cai for helpful discussion. Y.J.W., Y.Q.X., and H.A.N.L. acknowledge support from the 973 Program (2015CB857004), the National Natural Science Foundation of China (NSFC-11890693, 11421303), the CAS Frontier Science Key Research Program (QYZDJ-SSW-SLH006), and the K.C. Wong Education Foundation. M.F.G. acknowledges support from the National Science Foundation of China (grants 11873073 and U1531245). S.S.W. acknowledges support from the National Natural Science Foundation of China (NSFC-11673013).

\software{FTOOLS \citep[v6.21;][]{1995ASPC...77..367B},
HEAsoft \citep[v6.21;][]{2014ascl.soft08004N},
XSPEC \citep[v12.9.0;][]{1996ASPC..101...17A}}

\appendix
\section{Long-term light curves of the 14 sources in the total sample}


\setcounter{figure}{0}
\renewcommand{\thefigure}{A\arabic{figure}}

\begin{figure*}[!thbp]
\centering
\includegraphics[width=0.78\paperwidth, clip]{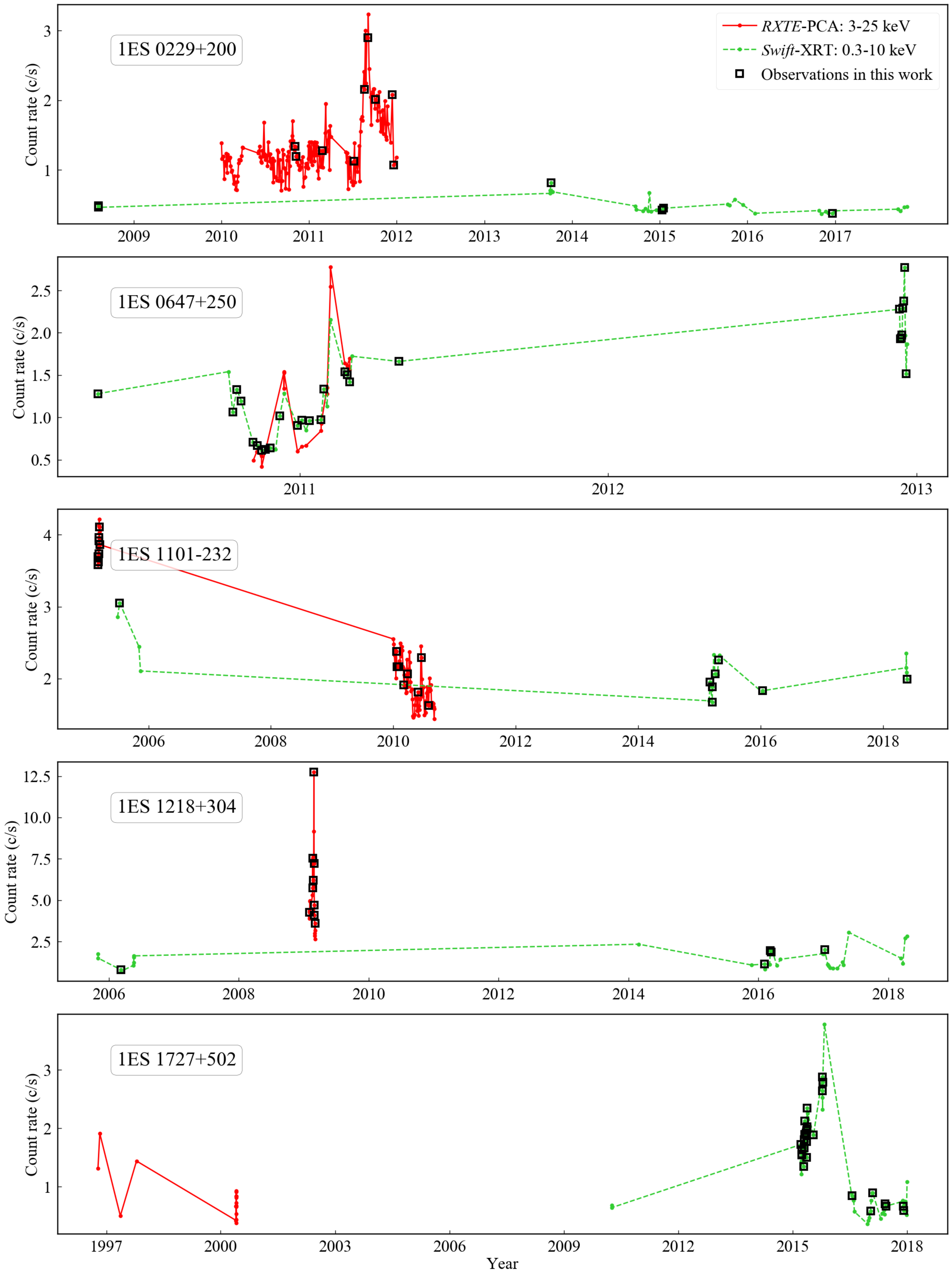}
\caption{Long-term light curves of the 14 sources in the total sample. The red segmented line represents the 3--25 keV \rxte/PCA light curve and the green dashed line represents the 0.3--10 keV \swift/XRT light curve. The black open squares denote the observations selected in this work. Note that for \rxte/PCA light curves, the units of y-axis are c s$^{-1}$ PCU$^{-1}$.}
\label{fig:lc}
\end{figure*}

\renewcommand{\thefigure}{\arabic{figure} (Cont.)}
\addtocounter{figure}{-1}

\begin{figure*}[!thbp]
\centering
\includegraphics[width=0.78\paperwidth, clip]{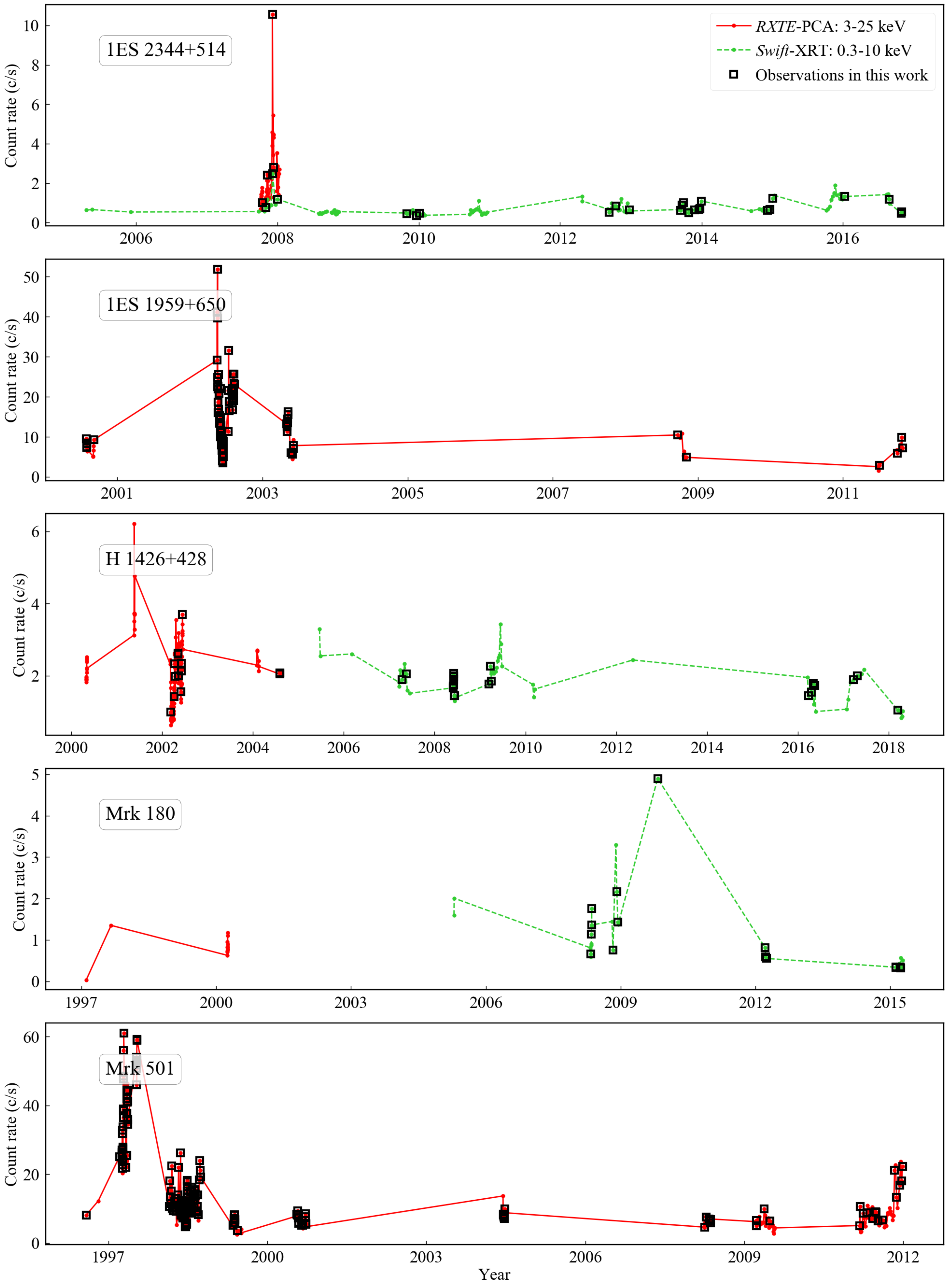}
\caption{}
\end{figure*}

\renewcommand{\thefigure}{\arabic{figure} (Cont.)}
\addtocounter{figure}{-1}

\begin{figure*}[!thbp]
\centering
\includegraphics[width=0.78\paperwidth, clip]{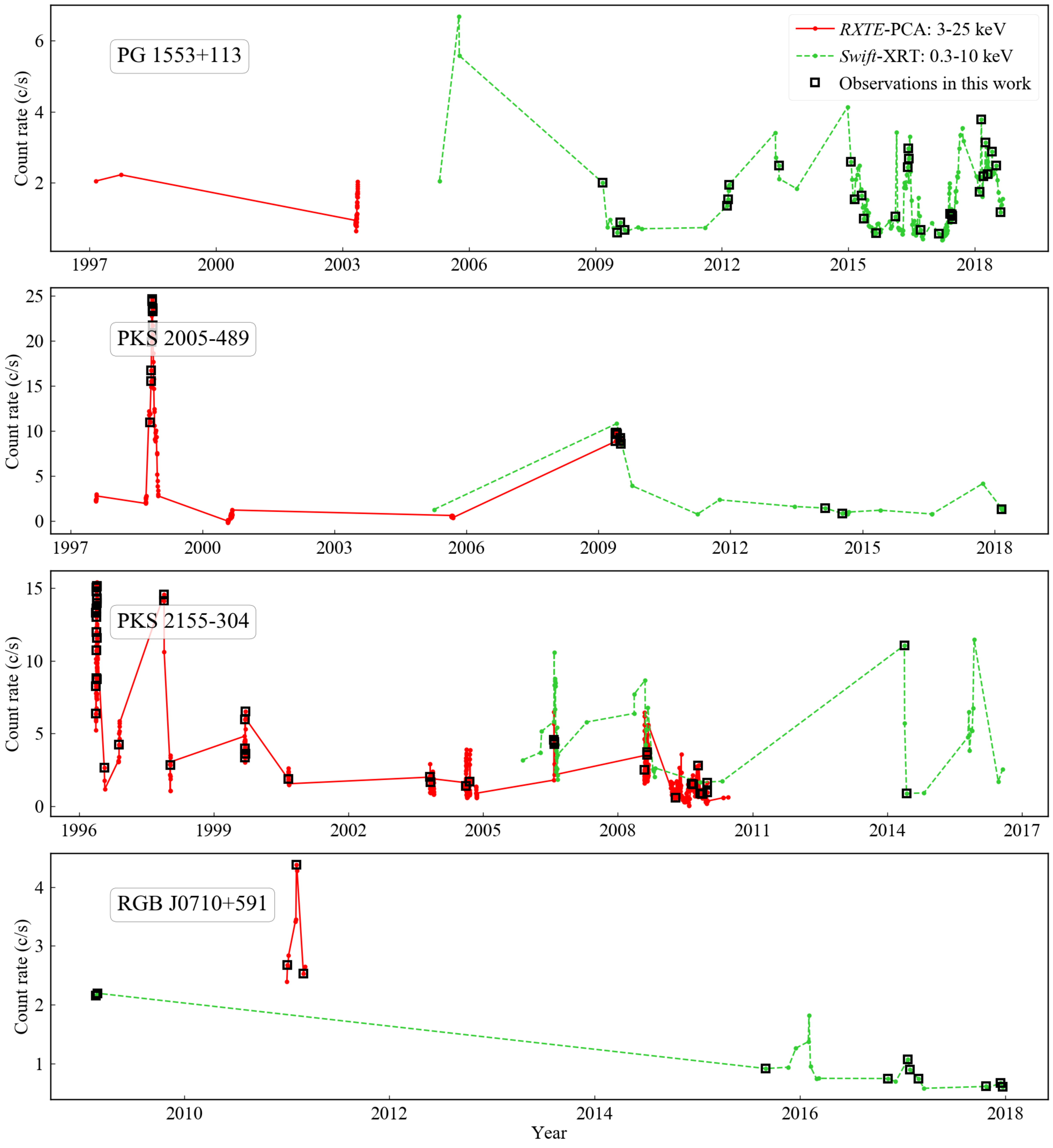}
\caption{}
\end{figure*}


\end{document}

%% file: table1.tex
1& 1ES 0229+200 & 0.140 & This work (9,6)$^\textrm{1,2}$, M08$^\textrm{3}$ & 8.92$\pm$0.70 & Wu09 \\
2& 1ES 0647+250 & 0.450 & This work (0,27)$^\textrm{2}$ & 7.73$\pm$0.70 & Wu09 \\ 
3& 1ES 1101$-$232 & 0.186 & This work (16,8)$^\textrm{1,2}$, M08$^\textrm{2,3,4}$ & \nodata & \nodata \\ 
4& 1ES 1218+304 & 0.182 & This work (9,8)$^\textrm{1,2}$, M08$^\textrm{2,3,4}$ & 8.47$\pm$0.70 & Wu09 \\ 
5& 1ES 1727+502 & 0.055 & This work (0,30)$^\textrm{2}$ & 7.84$\pm$0.15 & Woo05\\ 
6& 1ES 1959+650 & 0.048 & This work (80,0)$^\textrm{1}$, G06$^\textrm{1}$, M08$^\textrm{2,3,4}$, K18$^\textrm{2}$ & 8.14$\pm$0.70 & Wu09 \\ 
7& 1ES 2344+514 & 0.044 & This work (4,24)$^\textrm{1,2}$ & 8.57$\pm$0.70 & Wu09 \\ 
8& H 1426+428   & 0.129 & This work (14,22)$^\textrm{1,2}$, M08$^\textrm{2,4}$ & 8.51$\pm$0.70 & Wu09 \\ 
9& Mrk 180      & 0.045 & This work (0,16)$^\textrm{2}$, M08$^\textrm{2,3}$ & 8.10$\pm$0.70 & Wu09 \\ 
10& Mrk 501      & 0.034 & This work (227,0)$^\textrm{1}$, M08$^\textrm{2,3,4}$, K17$^\textrm{2}$ & 8.72$\pm$0.70 & Wu09 \\ 
11& PG 1553+113  & 0.500 & This work (0,31)$^\textrm{2}$, M08$^\textrm{2,3}$ & 7.25$\pm$0.70 & Wu09 \\ 
12& PKS 2005$-$489 & 0.071 & This work (24,4)$^\textrm{1,2}$, M08$^\textrm{3}$ & 8.57$\pm$0.14 & Wagner08 \\ 
13& PKS 2155$-$304 & 0.116 & This work (47,2)$^\textrm{1,2}$, M08$^\textrm{2,3,4}$ & 7.60$\pm$0.70 & Wu09 \\ 
14& RGB J0710+591 & 0.125 & This work (3,12)$^\textrm{1,2}$ & 8.67$\pm$0.70 & Wu09 \\ 

%% file: table2.tex
1& 1ES 0229+200 & 0.82 & < 0.001 & 1.08 $\pm$ 0.12 & $-$0.07 $\pm$ 0.09 & $-$0.17 & 0.533 & \nodata & \nodata \\ 
2& 1ES 0647+250 & 0.42 & 0.028 & 1.14 $\pm$ 1.33 & 2.03 $\pm$ 0.15 & $-$0.35 & 0.072 & \nodata & \nodata \\ 
3& 1ES 1101$-$232 & $-$0.49 & 0.014 & $-$0.33 $\pm$ 0.10 & 1.57 $\pm$ 0.04 & $-$0.87 & < 0.001 & $-$5.37 $\pm$ 0.72 & 4.23 $\pm$ 0.30 \\ 
4& 1ES 1218+304 & 0.90 & < 0.001 & 1.39 $\pm$ 0.10 & 0.75 $\pm$ 0.05 & $-$0.02 & 0.940 & \nodata & \nodata \\ 
5& 1ES 1727+502 & 0.47 & 0.009 & 1.28 $\pm$ 0.24 & $-$0.14 $\pm$ 0.04 & 0.05 & 0.777 & \nodata & \nodata \\ 
6& 1ES 1959+650 (2002) & 0.82 & < 0.001 & 0.95 $\pm$ 0.05 & 0.06 $\pm$ 0.04 & 0.43 & < 0.001 & 2.87 $\pm$ 0.23 & 0.61 $\pm$ 0.20 \\ 
& 1ES 1959+650 (2016) & 0.85 & < 0.001 & 0.60 $\pm$ 0.01 & 0.79 $\pm$ 0.00 & 0.51 & < 0.001 & 2.25 $\pm$ 0.14 & 2.18 $\pm$ 0.04 \\ 
7& 1ES 2344+514 & 0.67 & < 0.001 & 0.98 $\pm$ 0.09 & $-$0.59 $\pm$ 0.04 & $-$0.10 & 0.610 & \nodata & \nodata \\ 
8& H 1426+428 & 0.11 & 0.517 & \nodata & \nodata & $-$0.40 & 0.017 & $-$1.64 $\pm$ 0.24 & 3.25 $\pm$ 0.12 \\ 
9& Mrk 180 & 0.65 & 0.007 & 1.40 $\pm$ 0.17 & $-$0.13 $\pm$ 0.04 & 0.08 & 0.770 & \nodata & \nodata \\ 
10& Mrk 501 & 0.61 & < 0.001 & 0.89 $\pm$ 0.03 & $-$0.30 $\pm$ 0.02 & 0.60 & < 0.001 & 4.44 $\pm$ 0.15 & 0.17 $\pm$ 0.14 \\ 
11& PG 1553+113 & 0.23 & 0.207 & \nodata & \nodata & $-$0.46 & 0.008 & $-$3.80 $\pm$ 0.52 & 2.47 $\pm$ 0.07 \\ 
12& PKS 2005$-$489 & $-$0.03 & 0.888 & \nodata & \nodata & $-$0.42 & 0.026 & $-$2.55 $\pm$ 2.58 & 3.94 $\pm$ 0.42 \\ 
13& PKS 2155$-$304 & $-$0.75 & < 0.001 & $-$1.35 $\pm$ 0.13 & 1.70 $\pm$ 0.04 & $-$0.76 & < 0.001 & $-$3.49 $\pm$ 0.39 & 2.83 $\pm$ 0.12 \\ 
14& RGB J0710+591 & 0.62 & 0.014 & 1.08 $\pm$ 0.13 & 0.16 $\pm$ 0.07 & 0.24 & 0.398 & \nodata & \nodata \\ 